\documentclass[twocolumn]{aastex631}

\usepackage{enumitem}
\usepackage{longtable}

\begin{document}

\title{Comparing Optimized Systematic Error Correction Methods on Selected TESS Light Curves}

\correspondingauthor{David Rapetti}
\email{David.Rapetti@nasa.gov; drapetti@usra.edu}

\author[0000-0003-2196-6675]{David Rapetti}
\affiliation{NASA Ames Research Center, Moffett Field, CA 94035, USA}
\affiliation{Research Institute for Advanced Computer Science, Universities Space Research Association, Washington, DC 20024, USA}

\author[0000-0002-4715-9460]{Jon M. Jenkins}
\affiliation{NASA Ames Research Center, Moffett Field, CA 94035, USA}

\author[0000-0002-6778-7552]{Joseph D. Twicken}
\affiliation{NASA Ames Research Center, Moffett Field, CA 94035, USA}
\affiliation{SETI Institute, 339 Bernardo Ave., Suite 200, Mountain View, CA 94043, USA}

\author[0000-0003-1963-9616]{Douglas A. Caldwell}
\affiliation{NASA Ames Research Center, Moffett Field, CA 94035, USA}
\affiliation{SETI Institute, 339 Bernardo Ave., Suite 200, Mountain View, CA 94043, USA}

\author[0000-0002-6148-7903]{Jeffrey C. Smith}
\affiliation{NASA Ames Research Center, Moffett Field, CA 94035, USA}
\affiliation{SETI Institute, 339 Bernardo Ave., Suite 200, Mountain View, CA 94043, USA}




\begin{abstract}

The correction of systematic errors in TESS light curves is crucial for all astrophysical analyses employing these observations.~Here we present a data analysis and a software package, \textsf{SysCoCoPy}, to investigate and directly compare the performance of the Presearch Data Conditioning (PDC) correcting method from the Science Processing Operations Center (SPOC) pipeline and three correctors developed by the community. We incorporate these three correctors based on their implementations in \textsf{Lightkurve} and are particularly interested in the ability of the four correctors to remove scattered light contamination from the Earth and the Moon, which is a key systematic for TESS. We implemented these correctors in \textsf{SysCoCoPy} with a framework that allows an automatic optimization of their parameters to scale the analysis towards increasingly larger samples. \textsf{SysCoCoPy} provides qualitative and quantitative products for the comparison of individual cases as well as statistical results for selected samples. We currently find that while our automatic parameter optimization provides a significant number of successful scattered-light corrections for two of the correctors with a design that favors this purpose, an statistical analysis of our largest sample indicates that PDC is presently more robust, with a larger overall success level of the metrics used.

\end{abstract}

\keywords{\href{http://astrothesaurus.org/uat/498}{Exoplanets (498)}; \href{http://astrothesaurus.org/uat/1916}{Time series analysis (1916)}}


\section{Introduction} \label{sec:intro}

The Transiting Exoplanet Survey Satellite \citep[TESS;][]{Ricker2015JATIS} provides rich time series data to the community.~Accurate and precise removal of systematic errors from TESS light curves is not only critical for exoplanet investigations, but for all astrophysical research including stellar and innovative asteroseismology studies \citep[e.g.,][]{silvaAguirreRedGiants2020A,bowmanSeismology2020, wangSupernova2023}.
~Various approaches to correct for systematics such as scattered light from the Earth and the Moon, sky background light, instrumental effects including focus changes, pointing jitter of the spacecraft, and momentum dumps, while preserving intrinsic stellar variability, exist in the literature \citep[see e.g.][]{2012PASP..124.1000S, Stumpe2014, 2017MNRAS.471..759A, Luger_2016, Luger_2018, Nardiello:2019, Nardiello:2020, lundAsteroSystematics2021, hattoriUnpopular2022}.~However, no comprehensive analyses have been carried out to properly compare approaches and determine their validity on different types of targets in both a detailed and statistical manner.

Several approaches have been usually demonstrated on their own and on a limited set of selected targets that present advantages for those particular cases. There are also studies designed to apply specific methods more widely.~In order for the community to be able to apply and compare systematic corrections more generally, it is important to have an overall framework for directly and quantitatively comparing correctors as well as statistically examine their applicability on extended target samples. Here we present an initial study for this purpose and introduce the Python software \textsf{SysCoCoPy}\footnote{\url{https://github.com/drapetti/SysCoCoPy}}~\citep{Rapetti:24} to allow for both reproduction of the results presented and further analyses of other cases or samples of interest from the community.

Presearch Data Conditioning~\citep[PDC;][]{Stumpe2012,Stumpe2014,2012PASP..124.1000S} is the systematics correcting software module designed for and used in the Kepler pipeline \citep{2010ApJ...713L..87J}, and subsequently adapted for TESS and employed by the TESS Science Processing Operations Center~\citep[SPOC;][]{Jenkins2016SPIE}. PDC employs a Bayesian \textit{maximum a posteriori} (MAP) approach to systematic error correction in stellar light curves. In this study, we take as standard reference the PDC-corrected simple aperture photometry light curves as obtained by the TESS SPOC pipeline, and we compare them with those obtained using our implementations and automatic parameter optimizations of three systematic error correcting techniques developed by the community.~A specific, key objective of these comparisons is to investigate the relative ability of each corrector to properly remove scattered light contamination, which is a major systematic challenging to remove for TESS, affecting especially certain sectors and parts of the orbits\footnote{Data Release Notes (DRNs) are available for each TESS sector at \url{https://archive.stsci.edu/tess/tess\_drn.html}.}.

Building upon the widely-used Python software package \textsf{Lightkurve}\footnote{\url{https://docs.lightkurve.org}}~\citep{Lightkurve}, we have implemented in \textsf{SysCoCoPy} three systematic error correcting algorithms as described in Section~\ref{sec:methods}. \textsf{SysCoCoPy} is designed to carry out detailed comparisons of corrector performances between them and against PDC. The code allows the user to select a given set of 2-minute cadence targets (see Section~\ref{sec:flexible_targets} for other cadences) and sectors on which to apply the corrections.~We describe the samples used in this work in Section~\ref{sec:data}, and show representative examples of plots that qualitatively compare the performances between correctors in Section~\ref{sec:results}. These graphical representations are accompanied by metrics to provide quantitative comparisons.~We also obtain and show histogram compilations of metrics and discuss these statistical results in Section~\ref{sec:histograms}. We summarize our conclusions as well as planned analyses and future directions in Section~\ref{sec:future-work}.

\section{Methods} 
\label{sec:methods}

In this section, we describe the methods we use for the corrector performance analyses carried out in this work. First, in Section~\ref{sec:correctors} we list the correctors that we implemented in \textsf{SysCoCoPy} and used in the comparison analyses as well as several relevant characteristics of their implementation. Second, in Section~\ref{sec:comparing} we introduce the metrics employed to compare the correctors. Third, in Section~\ref{sec:optimization} we detail the corrector parameter values explored (see Table~\ref{tab:parametergrids}) as well as the automatic procedure and the metric that we use to optimize the set of values to be selected for each case and correction algorithm.

\subsection{Systematic error correctors} \label{sec:correctors}

The correcting algorithms currently implemented in \textsf{SysCoCoPy} and utilized for the comparison analyses presented in Section~\ref{sec:results} are:

\begin{enumerate}[label=(\alph*)]

\item\label{item:rcq} An adaptation of the systematics correction method from \cite{Hedges_2021} by using the \texttt{RegressionCorrector} class of \textsf{Lightkurve}. This algorithm employs (i) a spline polynomial fit aimed at accounting for stellar variability, (ii) Principal Component Analysis (PCA) eigenmodes to model the sky background light (see Section~\ref{sec:flux_adjustments} for details on background flux modeling), and (iii) the mean, standard deviation and skewness of the spacecraft quaternions\footnote{We currently used camera 1 for most of our cases, except for a few for which this induced a computational failure and camera 4 was used instead. We also noticed that the use of a different camera for cases without a failure might still interact differently enough with the spline. Both of these effects could be further studied and tested in future analyses.} to describe pointing jitter and mechanical momentum dumps (hereafter we refer to this correcting method as RCQ, standing for Regression Corrector with Quaternions).

\item\label{item:pld} An adoption of the Pixel Level Decorrelation (PLD) technique \citep{Deming_2009,Luger_2016, Luger_2018} as implemented in the \texttt{PLDCorrector} class of \textsf{Lightkurve}.~This corrector utilizes the same first two elements, (i) and (ii), of RCQ (although the parameter values might not be the same; see Table~\ref{tab:parametergrids}) for stellar variability and background light, but instead of quaternions for (iii), it uses PLD to account for pointing jitter and mechanical effects.~The PLD implementation in \textsf{Lightkurve} uses three different aperture masks.~First, the aperture mask used to create the Simple Aperture Photometry (SAP) light curve (see Section~\ref{sec:flux_errors} for details on SAP flux errors), for which we currently use a fixed parameter value of `pipeline'. This corresponds to the mask determined by SPOC. Second, the mask with the pixels used to create the background model, which should contain no flux from the target star as well as negligible flux from stars in the background\footnote{Note that in many postage stamps there can be hundreds to thousands of stars and might not be possible to choose background pixels that contain no flux from background stars.~The goal for this mask, however, is to select background pixels with insignificant flux levels from background stars.}.~For this, we employ by default the background mask obtained by SPOC. 
These pixels are used to generate the systematics model for the scattered and background light. Finally, an aperture mask with the pixels employed for the PLD basis vectors. As included in Table~\ref{tab:parametergrids}, for this mask we allow the parameter values of `pipeline' (i.e., from SPOC) and `threshold'.~The latter value consists of a mask with all the pixels brighter than three standard deviations above the median flux.~Note also that for TESS we use the default PLD order value of one, i.e., only the pixel fluxes, and not higher orders, are employed to construct the design matrix. See below for further details on our custom PLD parameters - otherwise, the rest also have default \textsf{Lightkurve} values.

\item\label{item:cbv} A version of PDC as adapted in the \texttt{CBVCorrector} class of \textsf{Lightkurve}.~This corrector utilizes the Cotrending Basis Vector (CBV) technique that the PDC method of the SPOC pipeline employs. Our implementation of this corrector in \textsf{SysCoCoPy} incorporates a spline polynomial fit, such as those used in the two correctors above.~This is added through an extra design matrix.~However, our current testing with the optimization procedure described in Section~\ref{sec:optimization} indicates that there are degeneracies between the spline and CBV parameters which advise us to not utilize the spline at this time\footnote{The use of the spline can be controlled by a flag, which should facilitate further investigations on the compatibility of these two elements if of interest for future work.} (see also below and Table~\ref{tab:parametergrids}). Hereafter we will refer to this corrector as CBV.

\end{enumerate}

Calibrated pixels and target star light curves are exported by SPOC with anomaly flags\footnote{SPOC data quality flags in archival target pixel and light curve files are defined in the TESS Science Data Products Description Document at \url{https://archive.stsci.edu/missions/tess/doc/EXP-TESS-ARC-ICD-TM-0014-Rev-F.pdf}.} that identify anomalous conditions associated with the instrument and/or data. The anomaly flags are provided for each target and cadence.~A bitmask may be specified in \textsf{Lightkurve} to filter out cadences exhibiting one or more types of anomalies. For the current results, we applied a quality bitmask of 17087 for all target stars and sectors. This mask excludes cadences with an attitude tweak (bit 1), safe mode (bit 2), spacecraft in coarse point (bit 3), spacecraft in Earth point (bit 4), argabrightening event (bit 5), reaction wheel desaturation event (bit 6), manual exclude due to an anomaly (bit 8), impulsive outlier removed before cotrending (bit 10), and bad calibration exclude (bit 15).

\subsubsection{Flux Adjustments}
\label{sec:flux_adjustments}

For PLD and RCQ, we desire to begin with calibrated pixels that are not background corrected, so we add the background flux estimated by SPOC into the calibrated and background-removed pixel values of the original target pixel file\footnote{Archival SPOC products including target pixel and light curve files are documented in the TESS Science Data Products Description Document at \url{https://archive.stsci.edu/missions/tess/doc/EXP-TESS-ARC-ICD-TM-0014-Rev-F.pdf}.} (TPF) before applying these correctors.~After the corrections, in order to preserve the true mean flux level, we apply the following flux adjustment to the corrected light curves.~We adjust the corrected flux by subtracting a constant level calculated as the third dimmest median pixel flux value times the number of pixels. This is inspired by the sky background bias correction applied by the SPOC pipeline, as documented in TESS DRN DR38\footnote{\url{https://archive.stsci.edu/missions/tess/doc/tess_drn/tess_sector_27_drn38_v02.pdf}.}.~We then adjust the flux for crowding by non-target stars and for the fraction of the target star flux captured in the photometric aperture using the methods of and the crowding and flux fraction values provided by the SPOC pipeline\footnote{In \textsf{SysCoCoPy}, the first flux adjustment is performed by the method \texttt{flux\_level\_adjust}, and the other two by the method \texttt{flux\_adjust}.}. We also use the flux fraction to scale the flux errors, but not the crowding since its effect on the flux errors is negligible.

For plots, such as those shown in Figures~\ref{fig:selected_cases_1}~and~\ref{fig:selected_cases_2}, we compare the corrected light curves for PLD, RCQ and CBV (in purple, blue, and green, respectively), as well as the PDC-corrected light curve from SPOC (in red).~The flux fraction and crowding adjustments are also applied to the reference SPOC SAP light curves \citep[in orange;][]{twicken:PA2010SPIE,morris:PA2020KDPH}. After all the flux adjustments, the metrics discussed in Section~\ref{sec:comparing} are computed for all the light curves in the figures and recorded in their top legend.

\subsubsection{Flux Errors}
\label{sec:flux_errors}

In \textsf{Lightkurve}, the flux errors produced by the \texttt{to\_lightcurve} method of the \texttt{tpf} class, using the \texttt{sap} option, do not match those in the SPOC SAP light curves, as downloaded from the Barbara A. Mikulski Archive for Space Telescopes\footnote{\url{https://mast.stsci.edu/}} (MAST). We traced this difference to the fact that this method assumes no correlation between the per pixel background estimates, while the SPOC pipeline employs a common background estimate for all pixels associated with each target star and cadence, and hence assumes full correlation\footnote{We explicitly confirmed that the difference in the flux errors of the light curves from these two sources can be accounted for by subtracting uncorrelated background flux errors from the flux errors obtained from the \texttt{to\_lightcurve} method, and adding fully correlated background flux errors, both in quadrature. A test method for updating the flux errors produced by the \texttt{to\_lightcurve} method in this manner is available in \textsf{SysCoCoPy}. However, since \texttt{to\_lightcurve} is employed as needed in \textsf{Lightkurve}, any changes that one might desire to apply to the level of correlation of background flux errors should be implemented in \texttt{to\_lightcurve} for consistency throughout.}. The \texttt{to\_lightcurve} method is used in the \texttt{PLDCorrector} class of \textsf{Lightkurve}.

In \textsf{SysCoCoPy}, for consistency we use \texttt{to\_lightcurve} to obtain the SAP light curve from the non-background-removed TPF before applying the RCQ corrector. When generating PLD and RCQ corrected light curves, in order to further ensure consistency with the flux errors calculated by \texttt{to\_lightcurve}, we proceed as follows. For calculating the over-fitting metric (see Section~\ref{sec:comparing}), we use a SAP light curve obtained by applying \texttt{to\_lightcurve} to the original SPOC TPF, instead of the corresponding SPOC SAP light curve.~The former SAP light curve is used as the before-correction light curve for the over-fitting metric, while the corrected light curves from PLD, RCQ, and CBV are employed as the after-correction light curves.~On the other hand, also for consistency, we use the SAP flux errors from SPOC, i.e., those accounting for full correlation between the background flux errors, when calculating the over-fitting metric of the PDC-corrected light curve from SPOC.

For the regressions of PLD and RCQ, we propagate uncertainties using the \texttt{propagate\_errors} flag in the \texttt{correct} method of the \texttt{RegressionCorrector} class of \textsf{Lightkurve}, which samples from a multivariate normal distribution of weights\footnote{The use of this flag required a bug fix in \textsf{Lightkurve}, which we provided via a pull request that was subsequently accepted.~Note also that this flag has not yet been implemented in the \texttt{CBVCorrector} class, although it has basic structure to allow so.}. For PLD and RCQ, to account for the addition of the background flux in the TPF, the background flux errors are also subtracted in quadrature before the corrections are performed, via the \texttt{prepare\_lc} method in \textsf{SysCoCoPy}.

\begin{table*}[ht!]
    \setlength{\tabcolsep}{4pt}
    \centering
    \begin{tabular}{cccc}
       \hline
       Parameter & PLD & RCQ & CBV \\
       \hline\hline
       Number of spline knots  & (6, 12, 24, 48) & (6, 12, 24, 48) & (0)\\
       PLD aperture mask & (pipeline, threshold) & N/A & N/A\\
       Number of PCA terms & (8, 16, 32) & (3, 6, 9) & N/A\\
       Added background & True & True & False\\
       CBV type & N/A & N/A & (MultiScale.1-2-3/Spike, SingleScale/Spike)\\
       CBV indices & N/A & N/A & (1-8/all, 1-8/all)\\
       CBV $\alpha$ regularization & N/A & N/A & (0.00001, 0.001, 1)\\
       \hline
    \end{tabular}
    \caption{Grid of selected parameter values for each corrector (PLD, RCQ, and CBV) search for optimization. See Section~\ref{sec:optimization} for further details on parameters selected to be fixed.}
    \label{tab:parametergrids}
\end{table*}

\subsection{Products for Corrector Comparisons}
\label{sec:products}

Our first goal is to examine the ability of RCQ and PLD to effectively remove  scattered light features in the time series data extracted from TESS images in the presence of instrumental systematics and a varied range of stellar properties and variability. For this, we produce the products described next. We also use these products to compare the RCQ and PLD results to those from the CBV corrector, for which at first order one can expect results more similar to those of PDC. These comparisons also include the reference SAP and PDC light curves produced by the SPOC pipeline.

For each case (i.e., a target star in a given sector) from the samples we selected for this work (see Section~\ref{sec:data}), our approach consists of producing (i) a series of metrics, calculated via algorithmic implementations in \textsf{Lightkurve}; and (ii) graphical representations\footnote{\textsf{SysCoCoPy} allows the user to choose between different display formats; we use `\texttt{joined\_panels}', in this paper.} of the corrected light curves for detailed visual inspections and investigations of individual cases when desired.~We present a few representative examples of (ii) in Figures~\ref{fig:selected_cases_1}~and~\ref{fig:selected_cases_2}, which include corresponding metrics of element (i) in the top legend.~In addition to these individual metric and graphical representations, we also produce statistical representations of metrics from our selected samples (see Section~\ref{sec:histograms}).

\subsubsection{Performance Metrics}
\label{sec:comparing}

In order to estimate the scatter remaining after correcting a light curve for systematic trends, the TESS SPOC pipeline computes a Combined Differential Photometric Precision (CDPP) metric that was developed for the Kepler pipeline and uses a wavelet-based algorithm to calculate the signal-to-noise ratio of a reference transit waveform on various timescales \citep{Christiansen_2012}.~The \texttt{estimate\_cdpp} method in \textsf{Lightkurve} employs instead a simpler CDPP proxy algorithm, discussed in \cite{Gilliland_2011} and \cite{Van_Cleve_2016}, that uses a Savitsky-Golay (sg) high-pass filter \citep{Savitzky_1964}.~We employ this sgCDPP measure with a window length five times the maximum time scale\footnote{Thus, the selected window length is of $5\times200=1000$ minutes. Longer durations could be explored at the expense of reducing the number of samples available for the sgCDPP calculation. We list such potential exploration in Section~\ref{sec:future-work}.}, for a range of reference durations:~30, 60, 120, 160, and 200 minutes (see e.g., the top legends of Figures~\ref{fig:selected_cases_1}~and~\ref{fig:selected_cases_2}).

To assess the level of introduced noise after applying a systematic error correcting algorithm, we employ the over-fitting metric implemented in the method \texttt{overfit\_metric\_lombscargle} of the module \texttt{metrics} in \textsf{Lightkurve}. This goodness-of-fit metric measures the broad-band power spectrum via a Lomb-Scargle periodogram and is calibrated such that a value of 0.5 means that the introduced noise due to over-fitting is at the same power level as the uncertainties in the light curve.

This is in fact similar to the over-fitting metric utilized and provided by the SPOC pipeline, which in combination with other metrics, helps guide the PDC optimization algorithm.
However, these two metrics and their calibrations are different enough for us to not expect matching results, and we instead also produce over-fitting metric values using \textsf{Lightkurve} for PDC to fairly compare them with those for the other correctors (see the right panel of Figure~\ref{fig:HM_OS_histograms}).

\subsection{Corrector Parameters Optimization}
\label{sec:optimization}

In this section, we describe how we extended our community corrector adaptations, presented in Section~\ref{sec:correctors}, to allow \textsf{SysCoCoPy} to automatically optimize corrector parameter values.~To motivate this extension, and for chronological context, in Section~\ref{sec:motivation} we briefly discuss initial corrector results obtained without this automatic optimization.~This input then led us to implement our current basic parameter optimization (see Section~\ref{sec:implementation}), which could be amplified in future work, as discussed in Section~\ref{sec:future-work}.

\subsubsection{Fixed Parameter Tests and Motivation}
\label{sec:motivation}

For our initial tests, we used mostly default corrector parameter values from \textsf{Lightkurve} and/or the literature. A number of cases showed the potential of RCQ to recover flux on cadences that had been masked in PDC due to high levels of scattered light from the Earth and/or the Moon.~PLD also demonstrated this capability in somewhat fewer cases in these initial tests with fixed, default values.~On the other hand, note that CBV utilizes by construction the same cadences as PDC. Overall, CBV appeared to provide corrections relatively closer to those of PDC, as to be expected.

After the initial selection of corrector parameter values, these first results involved neither manual nor automatic adjustments of the parameters. At this stage, we applied the correctors to relatively small sets of cases.~However, encouraged by these initial results of the three correctors, RCQ, PLD, and CBV, we implemented a multi-dimensional, coarse grid of selected parameter values aimed at automatically improving the performance of each corrector, when applying them to increasingly larger and more diverse samples.~This is the methodology that we employ in this paper, as described in the next section.

\subsubsection{First, Basic Implementation}
\label{sec:implementation}

For PLD, RCQ, and CBV, we choose to explore the parameters and their values shown in Table~\ref{tab:parametergrids}. We selected these specific parameters and values by running a series of tests with the sample of 24 cases chosen for presenting pathological scattered-light contamination, as found by the SPOC pipeline and PDC (see further details in Section~\ref{sec:data}).

For each corrector and case (defined as a target in a given sector), we minimize a goodness-of-fit metric to select the best set of parameter values from the options presented in Table~\ref{tab:parametergrids}. The metric chosen is currently the harmonic mean of the various sgCDPP durations stated in Section~\ref{sec:comparing} (hereafter, we refer to this metric as HM). The HM metric was selected to balance the post-correction photometric scatter at different time scales while penalizing outliers.

We find that this metric, in combination with the parameter values in Table~\ref{tab:parametergrids}, works reasonably well in our tests of the 24-case sample~\ref{item:scat}.~As discussed in the results section, however, this parameter grid is not optimized for types of stellar variability sufficiently different from the limited range represented in that sample.~This is also discussed in Section~\ref{sec:future-work}, in considering future extensions of this work towards applying \textsf{SysCoCoPy} more generally to larger samples. As shown in the results section, we also calculate, present, and statistically study the over-fitting metric described above, but at this time we do not use it to determine the best set of parameter values. Combining the HM with this over-fitting metric for complementarity would require a calibration study, as discussed for future work in Section~\ref{sec:future-work}.

For CBV, note that the CBV types and indices are not independent parameters (see further details in the corresponding \textsf{Lightkurve} documentation).~Also, within the \texttt{CBVCorrector} class, instead of the default \texttt{correct} method, which optimizes the correction by adjusting the L2-Norm regularization penalty parameter $\alpha$ based on goodness-of-fit metrics of over-fitting (see Section~\ref{sec:comparing}) and under-fitting (which measures mean residual correlation between the studied target and neighboring targets), we use the \texttt{correct\_gaussian\_prior} method, for which we vary $\alpha$ as indicated in Table~\ref{tab:parametergrids}.~Our initial testing, which included adjusting target scores, indicated that the \texttt{correct} method might not be stable enough to be employed in our automatic corrections without manually supervising each individual case. This method also increases computing time and requires information from neighboring targets to calculate the under-fitting metric.

Note also that as described in Appendix~\ref{app:diagnostic_plots}, we have implemented a diagnostic plot to examine our automatic selection of corrector parameter values.




\subsubsection{Corrected Light Curves FITS File Structure}
\label{sec:fitsfiles}

For this optimization framework, we extended the Flexible Image Transport System (FITS) file structure that \textsf{Lightkurve} uses to save light curves, as follows. For a given corrector, we store the corrected light curves for each multi-parameter variation set of values in, for example, Table~\ref{tab:parametergrids} into additional columns of the existing FITS file binary table extension.

In addition, we implemented another binary table extension in which we save the parameter exploration values themselves.~These sets of values are ordered according to the HM metric (see e.g., Figure~\ref{fig:diagnostic_plots}), which is stored in the first column of this binary table extension in ascending order.

\subsection{Extensions in a Separate Branch}
\label{sec:flexible_targets}

We have also extended \textsf{SysCoCoPy} capabilities to analyze 20-second cadence data, as well as 200-second cadence data as obtained from Full-Frame Images (FFI).~Both extensions include all the correctors above, and are currently incorporated in a separate branch named \textsf{feature/flexible\_targets}, which in the future might be merged.~The 200-second implementation allows to employ the High-Level Science Product (HLSP) data provided by SPOC to MAST, denoted as TESS-SPOC~\citep{caldwellFTL2020}.~For the 20-second implementation of the CBV corrector, we utilize the branch \textsf{20-sec\_CBVs\_PR} of a \textsf{Lightkurve} fork\footnote{\url{https://github.com/drapetti/lightkurve}} that includes a \textsf{Lightkurve} pull request\footnote{\url{https://github.com/jcsmithhere/lightkurve/tree/cbvcorrector_feature/load_20-sec_cbvs}} which enables the loading of 20-second CBVs from MAST within the \texttt{CBVCorrector}.

For users intending to apply \textsf{SysCoCoPy} on large samples, we also include in the repository template scripts to support parallelization. Our approach is based on the GNU command \texttt{parallel}.~We performed initial testing with these scripts at the NASA Advanced Supercomputing (NAS) facility, but these should be further tested and adapted to other systems by users as needed.

In addition, the \textsf{feature/flexible\_targets} branch includes code we developed to compare sgCDPP measurements on systematics-corrected light curves from the Quick-Look Pipeline\footnote{\url{https://tess.mit.edu/qlp/}}~\cite[QLP;][]{Huang2020TESS, Huang2020TESS2, Kunimoto:21, Kunimoto:22} and PDC, and generate corresponding plots. Examples of the latter are shown in Figure~\ref{fig:qlp_pdc_comparisons} of Appendix~\ref{app:qlp_pdc_comparisons} for a sample of 5,000 cases from Sector 56. QLP input for this analysis is from publicly available MAST data as an HLSP. QLP, at the TESS Science Office (TSO) in MIT, processes TESS FFIs by performing multi-aperture photometry to extract light curves for all targets captured in FFIs brighter than 13.5 TESS magnitude. We employ QLP light curves for which a correction method was applied to remove systematic errors without affecting the intrinsic stellar variability and are therefore directly comparable to PDC-corrected light curves from TESS-SPOC 200-second FFI data, which as noted above are also available from MAST as an HLSP.

\subsection{Stellar Variation Injections Code}

\textsf{SysCoCoPy} also incorporates now our Python software \textsf{InSimPy}, which is in its first stages of development. This code can inject simulations of stellar variability into TESS target pixel files (TPFs), to then study the ability of the PLD, RCQ, and CBV correctors to accurately and robustly preserve, in the corrected light curves, the injected stellar variability. InSimPy utilizes the publicly available Python code TESS PRF\footnote{\url{https://github.com/keatonb/TESS_PRF}}. The latter contains tools to display the TESS pixel response function (PRF) at any location of the detector, using the SPOC PRF included on the MAST portal, to ensure an accurate representation of the spatial stellar flux distribution as modulated by the pointing jitter time histories archived at MAST.

For initial testing purposes, we implemented basic injection one and two~\citep[inspired by][]{Fetherolf:2023} sinusoidal models and a fitting analysis, and tested different metrics to determine the ability of each corrector to preserve an injected stellar variation for given model parameter values. Such developments are available within \textsf{InSimPy} for future community studies as useful. Towards their application to larger case samples in future work, we implemented software tools to use publicly available files with TESS magnitude and CDPP information to select quiet stars to build testing samples.

\section{Test Samples} 
\label{sec:data}

In this work, we present and study results for cases in three samples. Each sample contains a set of selected TESS SPOC 2-min targets and sectors.~As indicated above, we denote as a distinct case, a given sector of a target.

\begin{figure*}[ht!]
\centering
\includegraphics[width=0.495\textwidth]{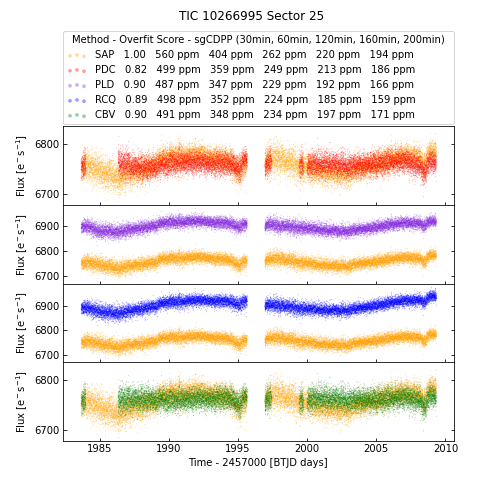}
\includegraphics[width=0.495\textwidth]{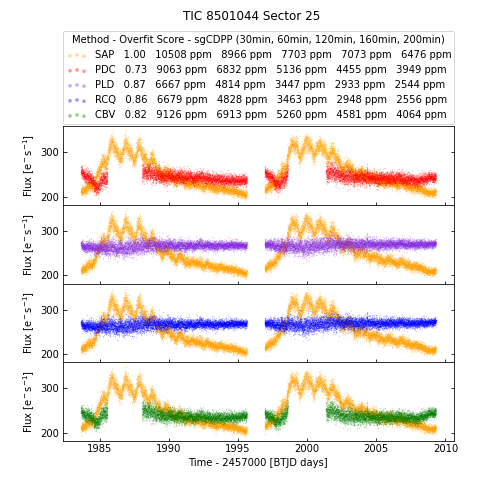}
\includegraphics[width=0.495\textwidth]{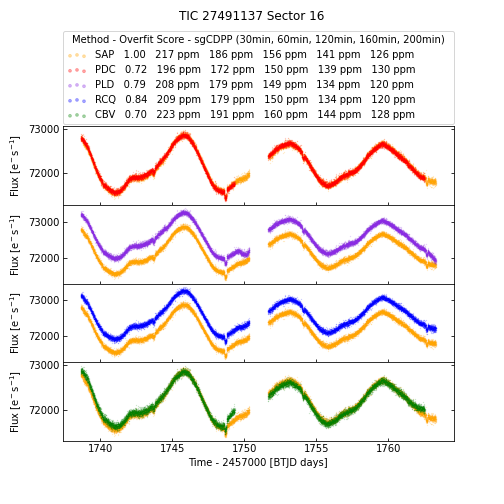}
\includegraphics[width=0.495\textwidth]{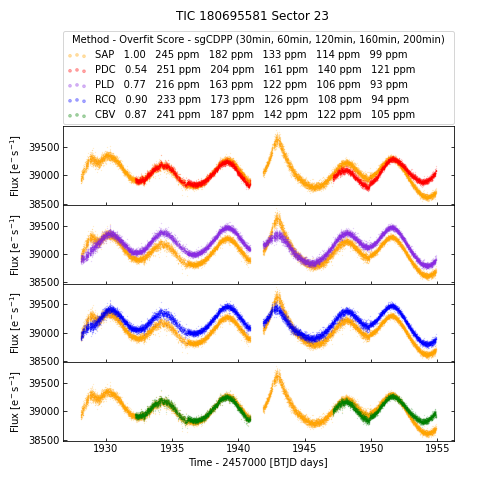}
\caption{Examples of successful systematic corrections for PLD (purple), RCQ (blue), and CBV (green).~The original SAP (orange) and PDC-corrected (red) light curves are also shown for reference and comparison purposes. Note that at the time of obtaining the data from MAST, the SAP and PDC curves in Sectors 14-26 were affected by a sky background bias which we correct for in PLD and RCQ as described in Section~\ref{sec:flux_adjustments}. Using our current parameter optimization, PLD and RCQ are able to smoothly recover significant portions of these light curves that were excluded by PDC due to scattered-light contamination. Top panels: two cases from sample~\ref{item:scat}, which contains cases with severe scattered-light contamination.~Bottom panels: two cases from sample~\ref{item:selec}; a rotationally variable star with three confirmed planets (TOI-2076; left panel), demonstrating the promise of such exploration for improving transit modeling;~and in the right panel, TOI-1807 exhibiting a light curve with large gaps in PDC at the start of the orbits, which are recovered by RCQ and PLD. The latter panel also indicates that the SPOC scattered-light flagging algorithm might be too conservative, excluding portions of the light curve that visual inspection shows as unaffected.~In the top legends, the second columns show the over-fitting score and the sgCDPP metrics are in parts per million (ppm).}
\label{fig:selected_cases_1}
\end{figure*}

\begin{enumerate}[label=(\roman*), wide, labelindent=0pt]

\item\label{item:scat} This sample consists of 24 cases selected for displaying levels and variations of scattered light capable of significantly contaminating the PDC results in cadences not excluded by this algorithm. The sample includes light curves where transit-like artifacts introduced when removing background scattered light triggered false positive Threshold Crossing Events (TCEs) in SPOC transit searches, and additional light curves from the same sectors representing a range of intrinsic and extrinsic variable stars.~We utilize this sample to study the ability of PLD and RCQ to account for this key systematic for TESS (see e.g., the two top panels of Figure~\ref{fig:selected_cases_1}). This sample contains a TESS Object of Interest (TOI)\footnote{\url{https://tess.mit.edu/toi-releases/}} corresponding to one case.

\item\label{item:selec} This somewhat larger sample contains 78 cases. Some of them were selected to include targets that exhibit a range of stellar variability including eclipsing binaries, pulsating variables, sinusoidal variables, and flare stars.~This sample also includes cases selected to test PLD, and four cases from \cite{Hedges_2021} to test RCQ. These cases contain TOIs, which together with other additional TOIs make a total of seven TOIs corresponding to 12 cases in the sample.~Finally, it also includes two cases selected for exhibiting systematic contamination likely attributable to asteroids passing through the respective postage stamps.

\item\label{item:catsel} The largest sample we have at this time was selected by applying a series of constraints on stellar properties in the MAST portal, and consists of 826 cases.~Rather than focusing on variable stars, this sample is intended to be more representative of the typical TESS target.~We used a range of $2717 < T_{\rm eff} < 10000$~K for the effective temperature, $4 < \log g < 5$ for surface gravity, and $7.98 < \rm{T_{mag}} < 8$ for TESS magnitude. The very narrow magnitude range was chosen to select a manageable number of targets for this test sample.

\end{enumerate}


For reproducibility purposes, these cases 
can be provided upon request via CSV (comma-separated values) files in the format that \textsf{SysCoCoPy} uses to read any given case sample to be analyzed.

\begin{figure*}[ht!]
\centering
\includegraphics[width=0.49\textwidth]{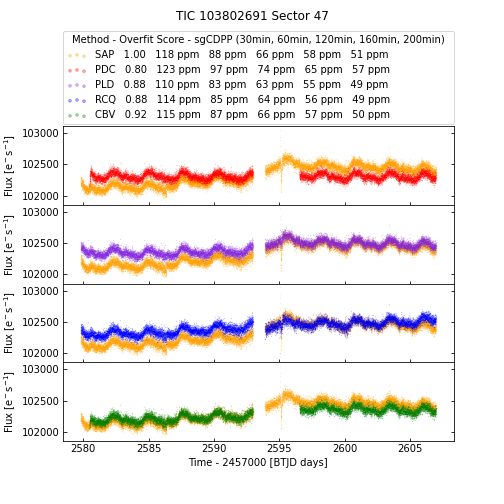}
\includegraphics[width=0.49\textwidth]{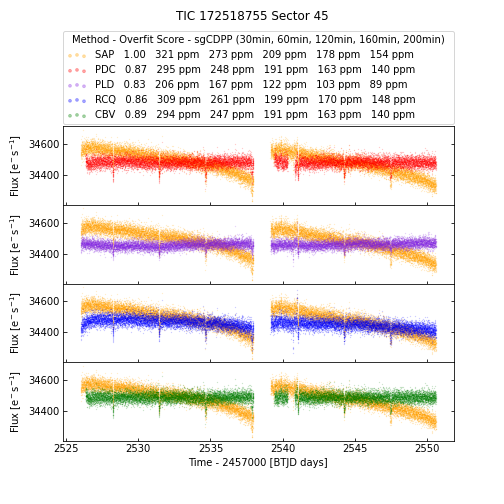}
\includegraphics[width=0.49\textwidth]{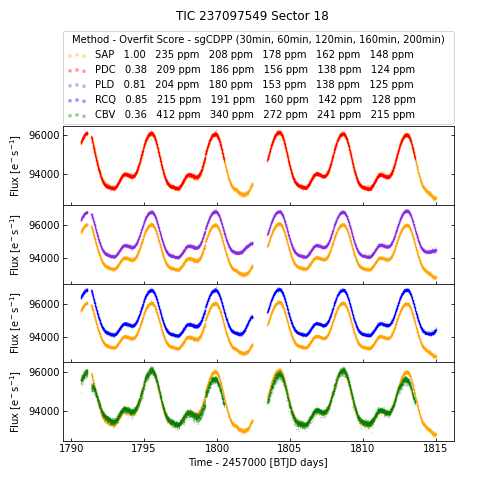}
\includegraphics[width=0.49\textwidth]{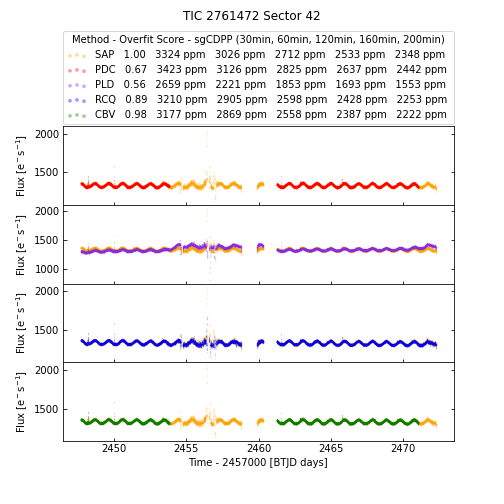}
\caption{Additional variety of successful examples in the same format as Figure~\ref{fig:selected_cases_1}, displaying similar PLD and RCQ recoveries of significant PDC-discarded portions of the light curves due to scattered light contamination. These four cases are from sample~\ref{item:selec}. The top, right panel shows TOI-1789, for which the transit at the beginning of the second orbit is better detected by PLD and RCQ.~The bottom, right panel, presents a particularly large gap at the end of the first orbit in PDC, which is relatively well recovered by RCQ and PLD.}
\label{fig:selected_cases_2}
\end{figure*}

\begin{figure*}[ht!]
\centering
\includegraphics[width=0.49\textwidth]{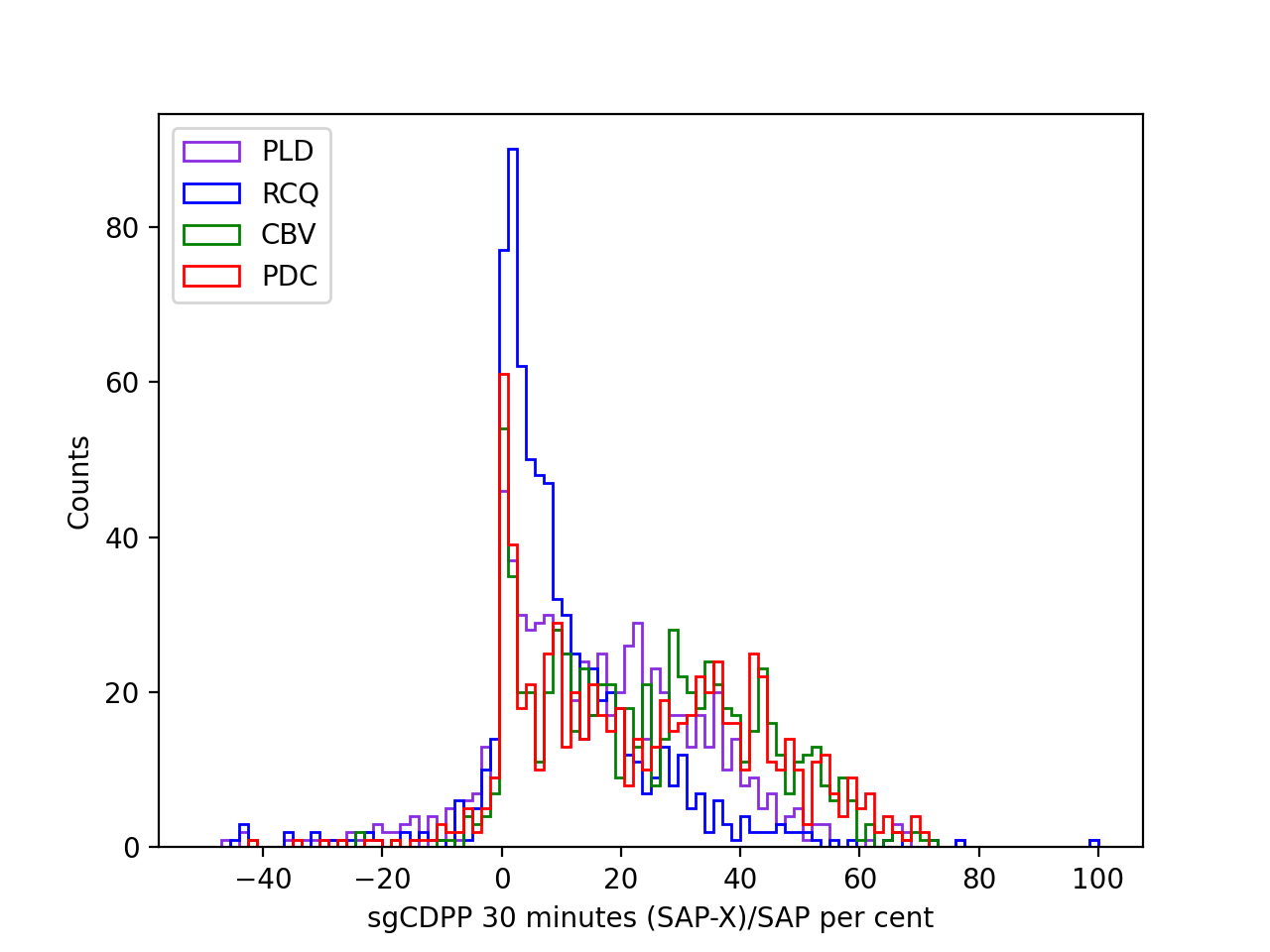}
\includegraphics[width=0.49\textwidth]{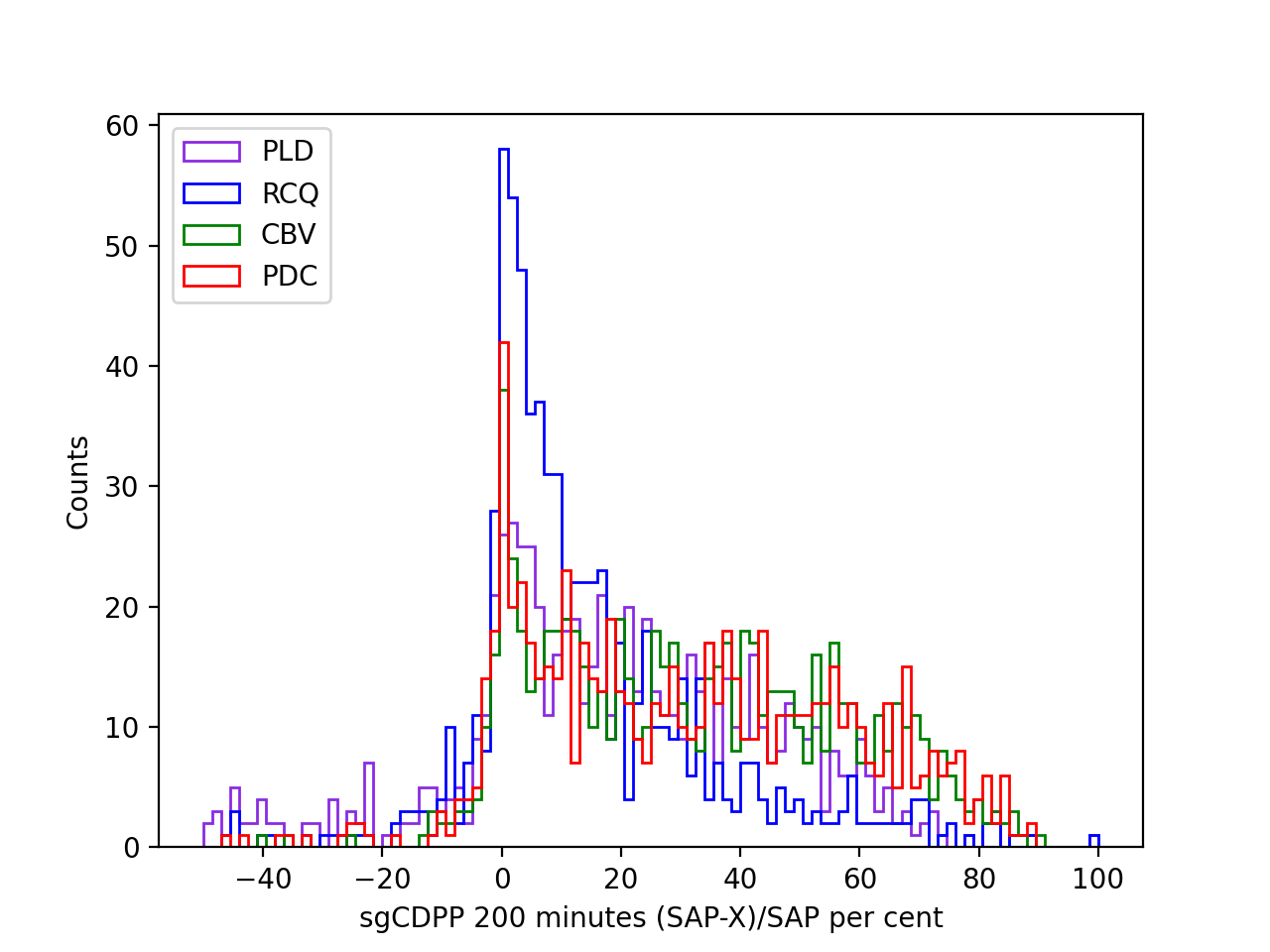}
\caption{For the largest sample analyzed~\ref{item:catsel}, these panels show the 1D histogram percentages of improvement for the sgCDPPs of the corrected light curves with respect to that of the SAP light curve, for the 30 (left) and 200 (right) minutes time scales. The color scheme for the correctors is the same as for the previous figures.
}
\label{fig:sgCDPPs_histograms}
\end{figure*}

\section{Results}
\label{sec:results}

In Figures~\ref{fig:selected_cases_1}~and~\ref{fig:selected_cases_2}, we show a total of eight representative cases from samples~\ref{item:scat} (the two top panels of Figure~\ref{fig:selected_cases_1})~and~\ref{item:selec} (the other six panels of the two figures). For all these cases, the corrections from PLD (in purple) and RCQ (in blue) allow the recovery of relatively significant portions of the light curves that were excluded by PDC (in red) due to scattered-light contamination in those cadences.~Such corrections thus have the potential to enhance exoplanet and stellar studies. Note in particular the cases with TOIs in the two bottom panels of Figure~\ref{fig:selected_cases_1} and the top right panel of Figure~\ref{fig:selected_cases_2}.

Encouraged by the promising results, we successfully applied our automatic PLD corrector parameter optimization to the TESS light curve of TOI-1835 in Sector 23, for \cite{Luque:23}. Remarkably, this paper discovered a resonant sextuplet of sub-Neptunes around the bright, nearby star HD 110067. The SPOC analysis, which includes a scattered-light flagging algorithm, that provided the PDC light curve in MAST for Sector 23, discarded cadences due to scattered light and high background contamination at the edges of the orbits. Using our implementation of PLD in \textsf{SysCoCoPy}, we were able to recover most of those cadences from the TESS data, allowing us to detect transits `b' (at the beginning of the first orbit) and `f' and `g' (at the beginning of the second orbit). All of them were found where predicted by their corresponding resonance. For both `f' and `g', this served to critically confirm their predicted 4:3 resonance by having a second detected transit for each.

Following the same primary motivation of recovering cadences discarded by PDC due to scattered light and high background contamination, we also applied our PLD and RCQ implementations to correct for systematics in the TESS light curve of TOI-1730 in Sector 60~\citep{Wilson:2026}. In this case, there was a smaller amount of data recoverable at the beginning of the first orbit, while the larger gap in the middle of the sector is due to data not collected because of a safe mode anomaly as well as data not acquired during the downlink preceding the anomaly. See also other applications of \textsf{SysCoCoPy} in works from \cite{Heidari:2024, von_stauffenberg:2024, Martioli:2024, Mireles:2026}.

Figures~\ref{fig:selected_cases_1}~and~\ref{fig:selected_cases_2} also include CBV corrected light curves (in green), which span the same cadences as PDC, and the uncorrected SAP light curves (in orange) for reference. For these cases, the figures show that CBV generally provides comparable results to PDC. As indicated above and shown in Table~\ref{tab:parametergrids}, we are currently not using the spline component with CBV. In testing the optimization framework with sample~\ref{item:scat}, various cases presented strong oscillations in the corrected light curves from CBV when using the spline. Inspection of the \texttt{CBVCorrector} class diagnostic plots reveals that the oscillations appear to be caused by the compensation between the spline and CBVs with similar time scales (mainly MultiScale bands 1 and 2, for these cases).
Without the spline, we tested and allowed the variations of the CBV parameters shown in Table~\ref{tab:parametergrids}: the CBV type together with its corresponding indices, and the CBV $\alpha$ regularization parameter\footnote{Future potential explorations could include a separate regularization with the spline, or a technique to project the spline onto the CBVs for orthogonalization purposes.} (for further details about the CBV parameters, see again the corresponding \textsf{Lightkurve} documentation and tutorial).



\begin{figure*}[ht!]
\centering
\includegraphics[width=0.49\textwidth]{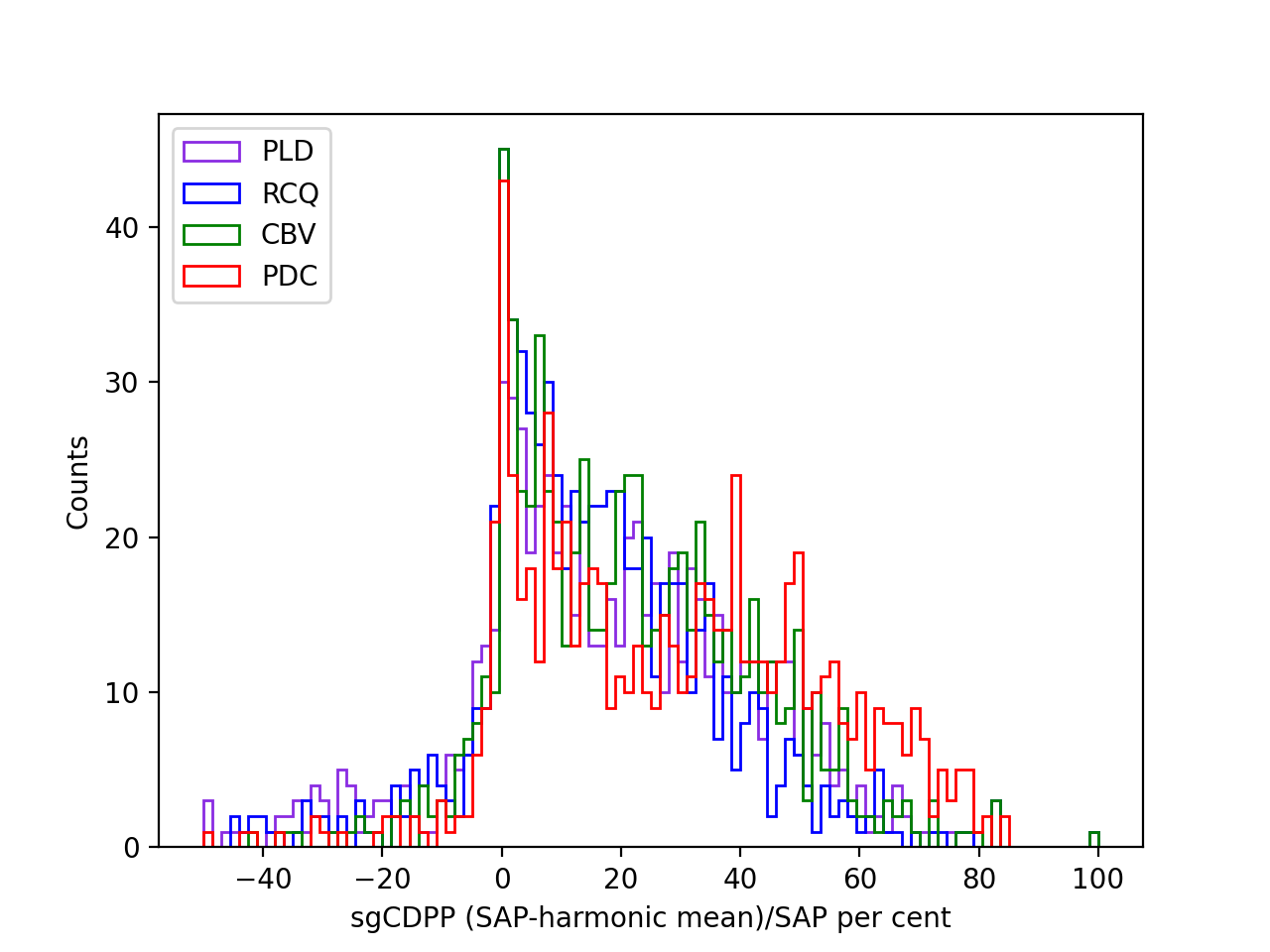}
\includegraphics[width=0.49\textwidth]{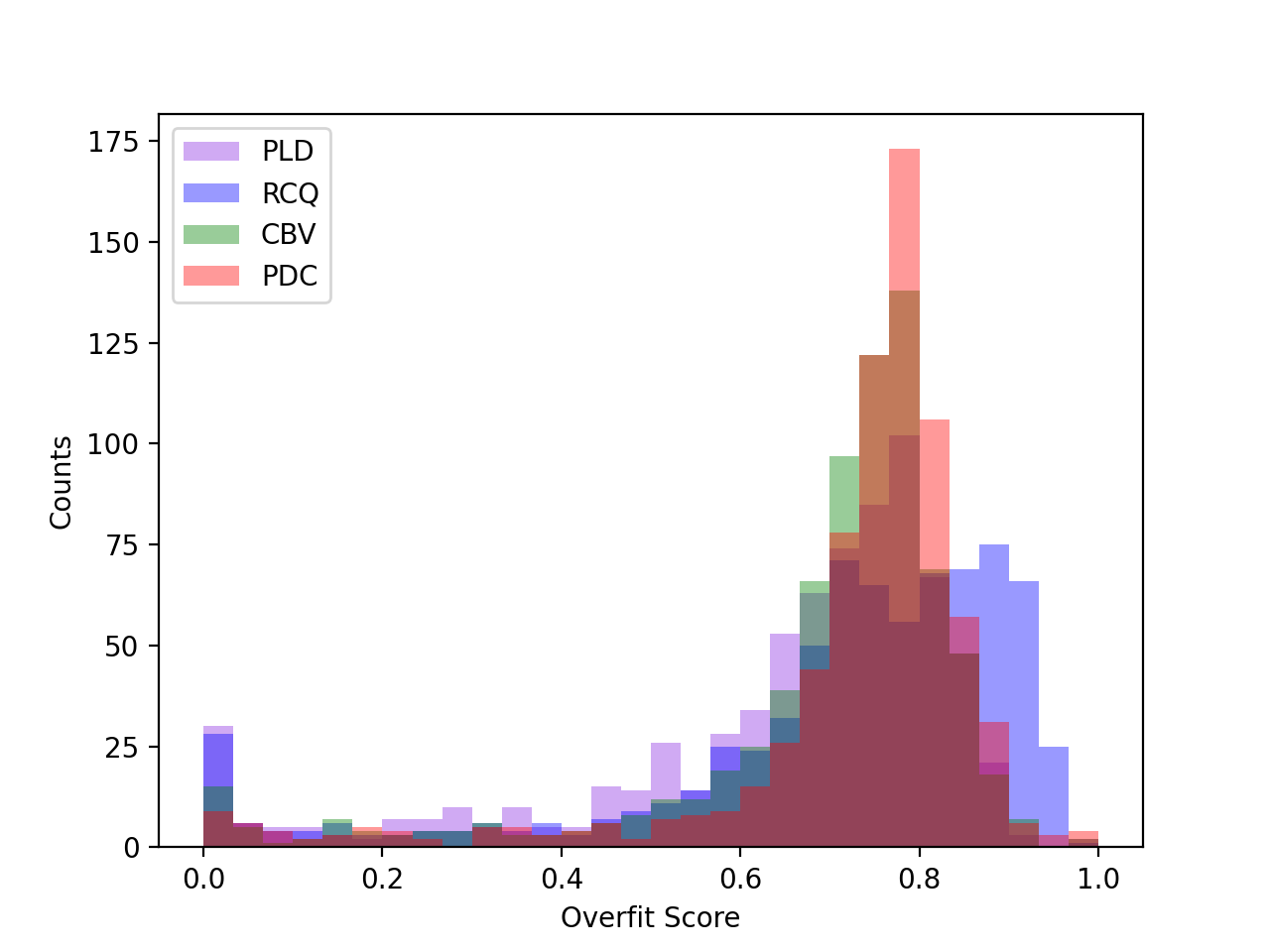}
\caption{For case sample~\ref{item:catsel}, the left panel shows the 1D histograms of the HM metric distribution in the same format as those in the panels of Figure~\ref{fig:sgCDPPs_histograms}. The right panel shows the over-fitting score distributions for each of the correctors using the same color scheme.}
\label{fig:HM_OS_histograms}
\end{figure*}

\subsection{Statistical Comparisons}
\label{sec:histograms}


In this section, we statistically analyze our largest case sample~\ref{item:catsel}, as follows.~Figure~\ref{fig:sgCDPPs_histograms} shows 1-dimensional (1D) histograms of an sgCDPP measure for each corrector and two of the duration times we employed, the smallest (30 minutes; left panel) and the largest (200 minutes; right panel). This measure (in the x-axis) is the percentage of the corrector sgCDPP distributions with respect to that of the SAP sgCDPP. We note the following points from these results:

\begin{itemize}

\item For all correctors (using the same color scheme as before), this sgCDPP measure has a peak around 0\%, with a broad positive tail, indicating targets for which the sgCDPP was lower for the corrected light curves. The negative side of the peak is sharp with a long low-level tail (truncated in the figure to better display the rest of the histogram). This tail contains the small number of cases where the corrected light curves exhibit larger sgCDPPs than the uncorrected (SAP) light curves.

\item RCQ has the largest peak around 0\% improvement and the smallest positive tail of successes (improved sgCDPP), followed by PLD. Interestingly, CBV and PDC present similar relatively bimodal shapes with a second, smaller peak around sgCDPP improvements of 30-40\%.

\item Given the somewhat opposite behavior of these distributions with respect to those for the over-fitting score (see the right panel of Figure~\ref{fig:HM_OS_histograms}), it is useful to examine the correlation between the two measures in these cases.\footnote{Note that \textsf{SysCoCoPy} can also provide 2-dimensional histograms.
}~For the 30 minutes time scale, the absolute value of the Pearson correlation coefficient, $\rho$,
~is higher for PLD ($\rho=0.14$) and RCQ ($\rho=0.11$) than for CBV ($\rho=-0.05$) and PDC ($\rho=-0.07$).~This difference is more significant for the 200 minutes time scale.~For the latter, the absolute value of the coefficients for PLD ($\rho=0.18$) and RCQ ($\rho=0.14$) are a bit higher, and those for CBV ($\rho=0.00$) and PDC ($\rho=0.02$) a bit lower than those for the 30 minutes time scale.~For the 200 minutes scale, the correlation between the sgCDPP and over-fitting metrics is relatively negligible for CBV and PDC.

\item Comparing the left and right panels of Figure~\ref{fig:sgCDPPs_histograms}, the sgCDPP measure distributions widen with the larger duration time (200 with respect to 30 minutes) on both directions i.e., towards the ‘error-augmentation’ negative side as well as the ‘error-reduction’ positive side.~On visual inspection, while for the 30 minutes time scale most cases have sgCDPPs below about a 50\% improvement for PLD and RCQ and about a 70\% for CBV and PDC, for 200 minutes the majority are below about an 80\% improvement for PLD and RCQ and about an 85\% for CBV and PDC.

\item As discussed above, SPOC uses a different CDPP metric. We compared the results from SPOC for an available time scale of 60 minutes to those from the sgCDPP and found a somewhat different distribution. The SPOC-PDC CDPP distribution is unimodal, without a peak around 0\%, and with a peak around 30\% that is relatively similar to that of the second mode of the sgCDPP distribution. It is, however, mostly symmetric on both sides, without either a sharp transition around 0\% or a long negative tail, and it dives off smoothly all the way towards 100\%. This comparison and the potential use of the SPOC CDPP is an area for future investigation, as discussed in Section~\ref{sec:future-work}.

\end{itemize}

Following the same format used in the panels of Figure~\ref{fig:sgCDPPs_histograms}, the left panel of Figure~\ref{fig:HM_OS_histograms} displays 1D histograms for the sgCDPP improvement of the HM metric, which combines all our time scales. The right panel of this figure shows the corresponding over-fitting 1D histograms for each of the correctors. Note the following points for this figure:

\begin{itemize}

\item In the left panel, the RCQ histogram does not present anymore the highest peak. In addition, PLD, RCQ and CBV appear to perform more comparably to each other based on these HM metric sgCDPP distributions, and PDC statistically outperforms all of them, with a second mode peak around 40\% improvement.

\item The over-fitting score distributions for each of the correctors appear relatively consistent with a peak around 0.8 and a longer tail downwards, except for RCQ that presents a bimodality centered around 0.8 with a second mode around 0.9. Within this second mode, RCQ thus presents more cases with less over-fitting than the other correctors.

\item Similarly to the sgCDPP case, the over-fitting metric implemented in \textsf{Lightkurve} is not fully equivalent to that used by SPOC, and more importantly the calibrations are different.~In this case, the distributions of the two metrics appear significantly different from each other, which should be explained in a major part by the calibration, and at a minor level by the implementation, as well as potentially by the different flux errors of the uncorrected light curves (see  Section~\ref{sec:flux_errors}), which are used as input for the calculation of the two metrics.

\end{itemize}

\subsection{Testing Stellar Variation Preservation}

Furthermore, we plan to develop simulated studies with which to test the performance of correctors via controlled, realistic injections of flux variations.~These simulations should allow us to investigate the ability of the correctors to properly preserve stellar variability.

\section{Conclusions and future work}
\label{sec:future-work}

To facilitate the key task of accurately correcting for systematic errors in TESS light curves in a variety of cases, here we presented a methodology to compare correctors that provides individual graphical and metric evaluations as well as statistical comparisons for samples of cases. This methodology also allows for the automatic optimization from a given set of corrector parameters based on the minimization of the harmonic mean of multiple sgCDPP metrics at different time scales. We implemented these comparison analysis tools into the Python software \textsf{SysCoCoPy}, to be made publicly available upon publication, and which builds upon the community package \textsf{Lightkurve}.

In particular, we have currently implemented the RCQ~\ref{item:rcq} and PLD~\ref{item:pld} correctors to especially investigate scattered light contamination for pathological cases as identified by SPOC via PDC -- see for instance sample~\ref{item:scat}. We have also incorporated the CBV~\ref{item:cbv} corrector which is inspired by the CBV technique implemented in PDC. The parameters for PLD, RCQ and CBV were optimized as described above, using the relatively extensive grid of parameter values shown in Table~\ref{tab:parametergrids}.~Using this discrete grid and the HM metric, we already obtained remarkable results for a number of cases from selected samples~\ref{item:scat}~and~\ref{item:selec}~(see Figures~\ref{fig:selected_cases_1}~and~\ref{fig:selected_cases_2}). The capability of this analysis was also demonstrated in~\cite{Luque:23}.~Applying \textsf{SysCoCoPy} to the study of a resonant multi-planetary system importantly supported the identification of transits previously missed in the TESS light curve due to scattered light contamination.

Note, however, that from visual inspection, we also find that for a non-negligible number of cases, for PLD and RCQ the present automatic optimization is not enough to avoid evident failures that could likely be resolved either manually (which is not feasible for large samples) or by amplifying the grid of parameter values to account e.g., for targets with higher stellar variability.~Such an amplification though should naturally increase computation time.

In addition to analyzing individual cases, we also analyzed our largest sample~\ref{item:catsel} from a statistical point of view.~This sample was selected by constraining stellar properties at the level needed to reach a sample about an order of magnitude larger than~\ref{item:selec}.~Based on 1D histograms of the improvements of sgCDPP in corrected light curves with respect to their uncorrected counterparts (see e.g., Figure~\ref{fig:sgCDPPs_histograms}), we find that PDC produces a larger number of cases with a greater level of sgCDPP improvement. This is also the case for the HM metric (see the right panel of Figure~\ref{fig:HM_OS_histograms}), which we employ for parameter optimization. Note also that CBV produces similar distributions to PDC for the 30 and 200 minutes sgCDPP metrics (Figure~\ref{fig:sgCDPPs_histograms}), but for the HM metric its distribution resembles more those of PLD and RCQ (right panel of Figure~\ref{fig:HM_OS_histograms}).

It is important to point out, however, that by visual inspections again, we find individual cases for which the sgCDPP metrics, and thus the HM metric, fail to identify invalid corrections, and that interestingly sgCDPP was not specifically designed for these purposes; hence, other metrics could also be considered and implemented for enhancements, within the same framework.~These cases might even contribute spurious counts of improvement to the histograms of PLD and/or RCQ.~In these cases, either PLD, RCQ, or both, produce light curves that display uncorrected scattered-light contamination in the cadences removed by PDC due to this systematic.~These can be identified clearly using the corrector comparison figures generated by \textsf{SysCoCoPy}, but the sgCDPP metrics might not register them as correction failures and might even evaluate them as error reductions.~According to this qualitative observation, the actual difference in successful corrections between PDC and PLD/RCQ might be more statistically pronounced that the sgCDPP histograms of Figures~\ref{fig:sgCDPPs_histograms}~and~\ref{fig:HM_OS_histograms} reveal. To improve upon the evaluation of PLD and RCQ performance, as well as the other correctors, metrics to be employed could be further investigated.

The potential use of SPOC CDPP might provide distributions without the asymmetric peaks around 0\% improvement shown in Figures~\ref{fig:sgCDPPs_histograms}~and~\ref{fig:HM_OS_histograms}.
~Another possibility would be to study adding sgCDPPs with longer durations to capture the performance at these time scales. Also, a metric for parameter optimization that combines the HM sgCDPP metric on one side, to function as a goodness-of-fit, and the over-fitting score on the other side, to function as a penalty term, could be explored. For this combination, the weights of the two metrics should be calibrated.~Alternatively, instead of combining these metrics, the over-fitting score could be used as a so-called `satisficing' metric, while keeping the HM as the `optimizing' metric.

In summary, \textsf{SysCoCoPy} can be utilized and extended by the community as useful to correct TESS light curves for systematics using and comparing different algorithms, while optimizing their parameters, and study the results either individually or statistically in samples.

\begin{acknowledgments}
We are grateful to Evan Tey and Chelsea Huang for sharing with us Python code to read and utilize TESS quaternion data,  as well as to Christina Hedges for her guidance in the first stages of implementing RCQ, and to Ben Tofflemire for input in recent stages of \textsf{InSimPy}. 
We acknowledge the use of public TESS data from pipelines at the TESS Science Office and at the TESS Science Processing Operations Center.~DR was supported by NASA under award number NNA16BD14C for NASA Academic Mission Services, under cooperative agreement award number 80NSSC25M7110, and under grant award number 80NSSC23K0439. Resources supporting this work were provided by the NASA High-End Computing (HEC) Program through the NASA Advanced Supercomputing (NAS) Division at Ames Research Center for the production of the SPOC data products, as well as for data analyses presented in this work. This research made use of \textsf{Lightkurve}, a Python package for Kepler and TESS data analysis~\citep{Lightkurve}. This work made use of Astropy:~a community-developed core Python package and an ecosystem of tools and resources for astronomy \citep{astropy:2013, astropy:2018, astropy:2022}. The data analyzed in this paper were obtained from MAST at the Space Telescope Science Institute (STScI). 
STScI is operated by the Association of Universities for Research in Astronomy, Inc., under NASA contract NAS5–26555. Support to MAST for these data is provided by the NASA Office of Space Science via grant NAG5–7584 and by other grants and contracts.
\end{acknowledgments}

%

\vspace{5mm}
\facilities{TESS, MAST}


\software{Lightkurve \citep{Lightkurve}, Astropy\footnote{\url{https://www.astropy.org}} \citep{astropy:2013, astropy:2018, astropy:2022}.}



\appendix

\begin{figure*}[ht!]
\centering
\includegraphics[width=0.495\textwidth]{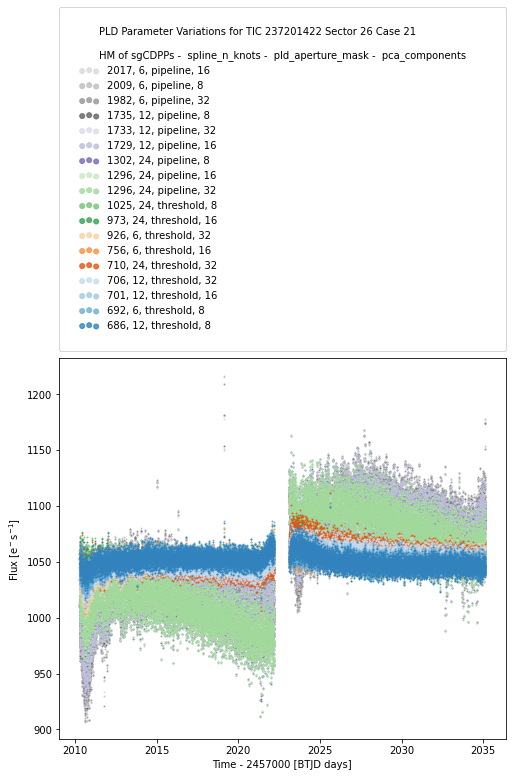}
\includegraphics[width=0.495\textwidth]{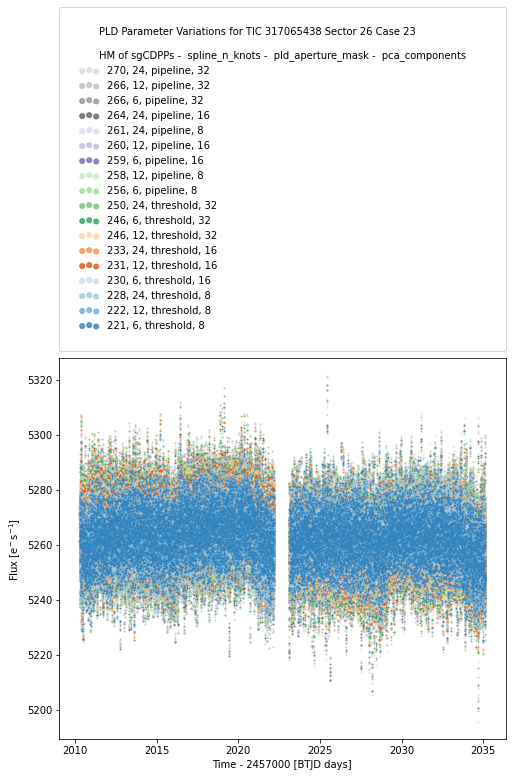}
\caption{Two examples of parameter variation diagnostic plots for PLD. The legend shows the harmonic mean (HM) of the sgCDPPs in ppm for our various time scales and a set of parameter combinations, in decreasing order of the HM metric.}
\label{fig:diagnostic_plots}
\end{figure*}

\section{Optimization Diagnostic Plot}
\label{app:diagnostic_plots}



We have implemented an optional diagnostic plot for the user to inspect the corrected light curves comparing them with the HM metric for a chosen set of corrector parameter values. Note also that a different flag is available to trigger standard diagnostic plots from \textsf{Lightkurve} for each chosen corrector. This is to allow further inspection of an individual correction by examining the separate components of the fit. See the \textsf{Lightkurve} documentation on the corrector of interest for further details.

Figure~\ref{fig:diagnostic_plots} presents a couple of representative instances to show the format of the diagnostic plot that is available in \textsf{SysCoCoPy} to inspect the light curves produced by a set of corrector parameter variations. The first column of the top legend displays the HM metric that is used to order these parameter variations (the rest of the columns) and select the set that minimizes it (at the bottom of the legend) and thus the corresponding light curve (in blue) as the one to be compared with the results from the other correctors in figures like~\ref{fig:selected_cases_1}~and~\ref{fig:selected_cases_2}.

For these PLD cases, the HM improvements within the range of selected parameter variations (note that for convenience this is a smaller set that the one presented in Table~\ref{tab:parametergrids}) are about 66 (left panel) and 18 (right panel) per cent. These cases were manually chosen to roughly represent a more extreme case and a more typical one, respectively, based on a relatively small sample selection.

\begin{figure*}[ht!]
\centering
\includegraphics[width=0.495\textwidth]{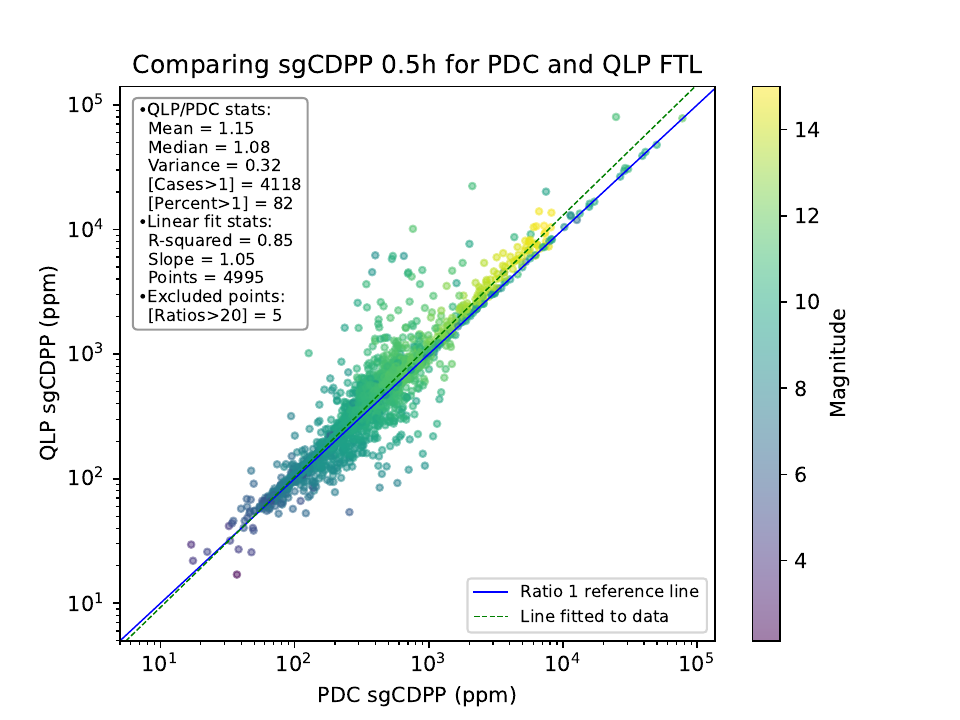}
\includegraphics[width=0.495\textwidth]{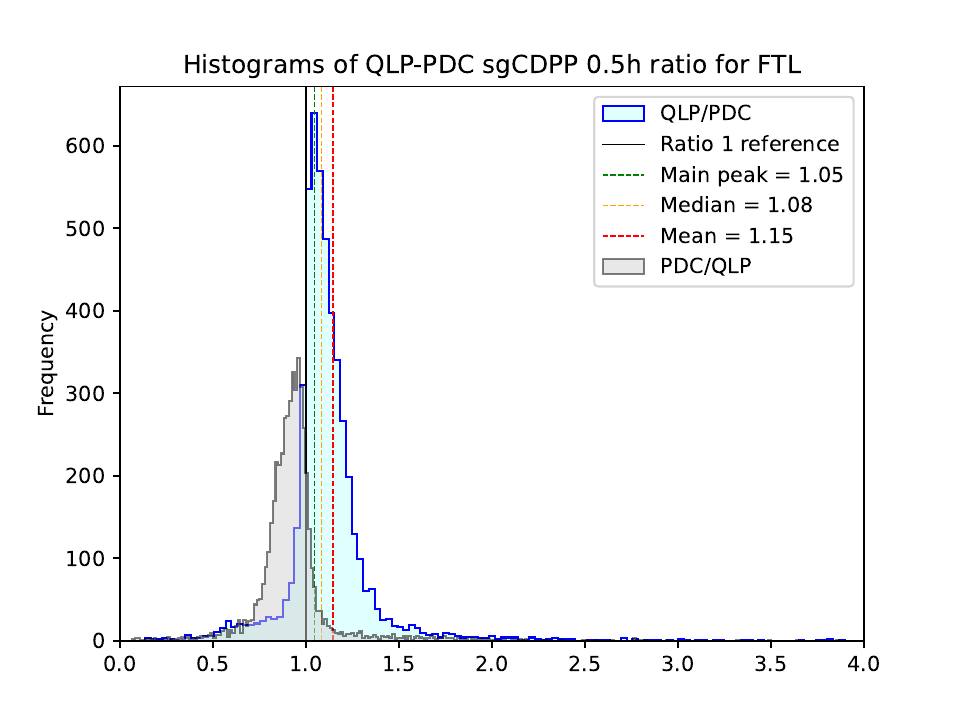}
\includegraphics[width=0.495\textwidth]{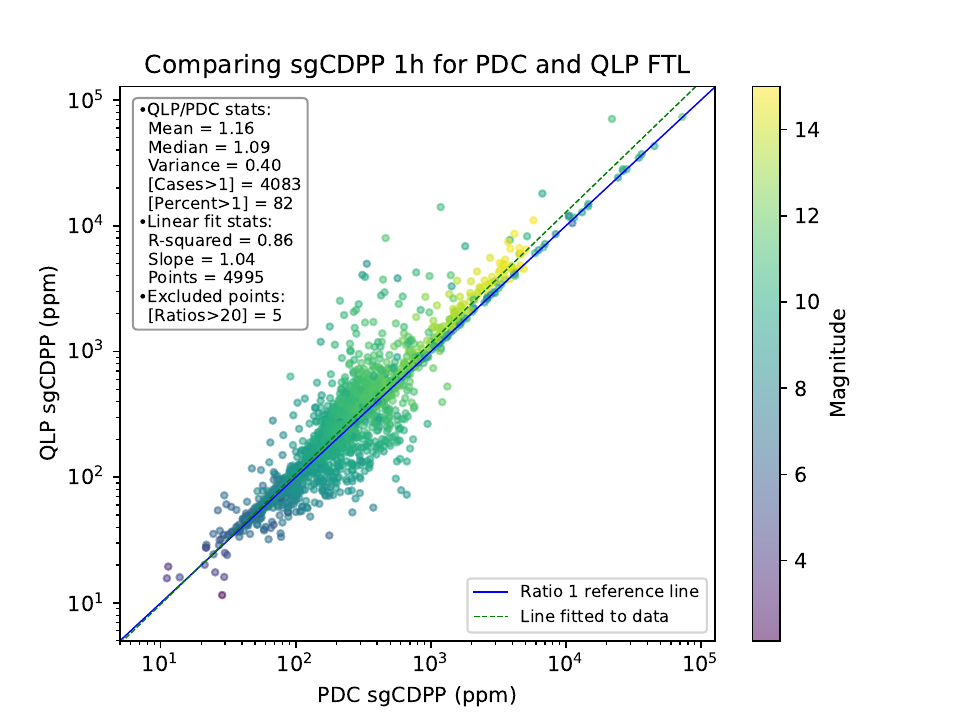}
\includegraphics[width=0.495\textwidth]{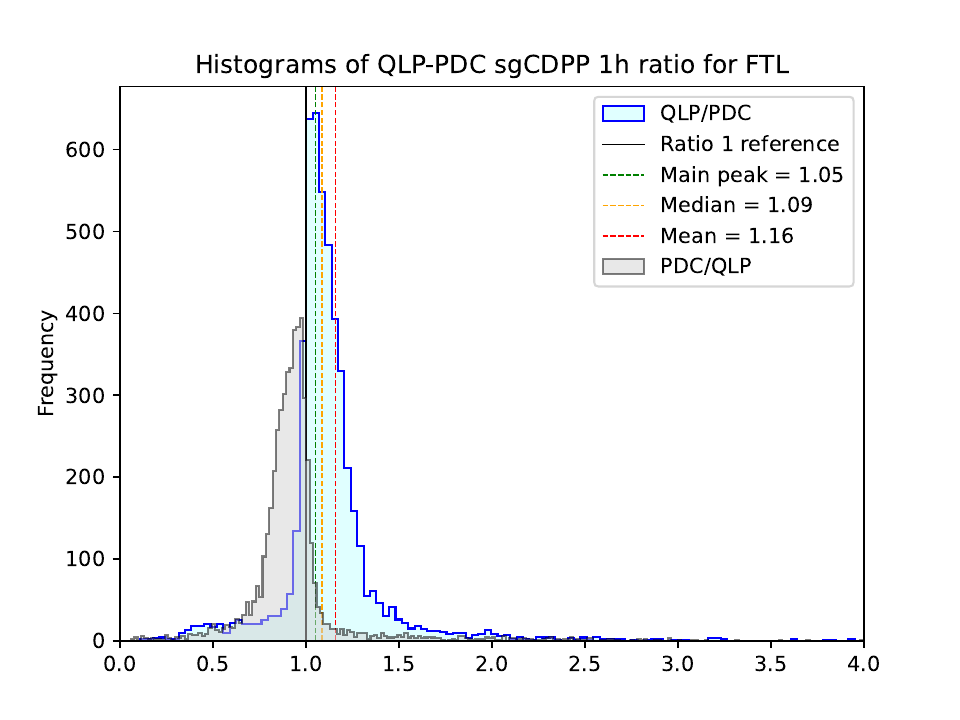}
\includegraphics[width=0.495\textwidth]{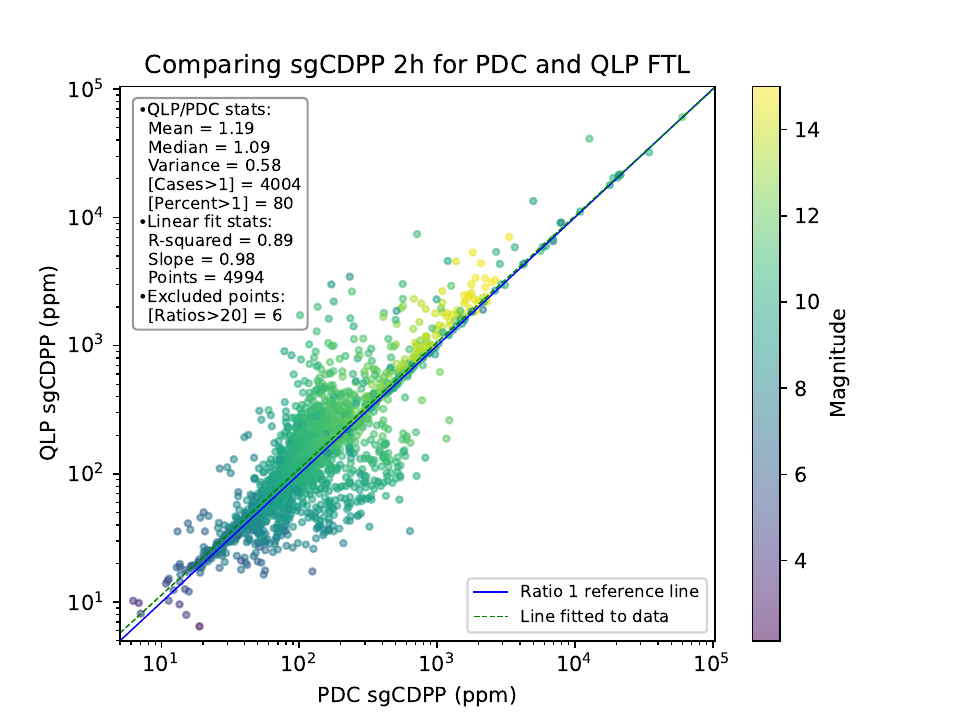}
\includegraphics[width=0.495\textwidth]{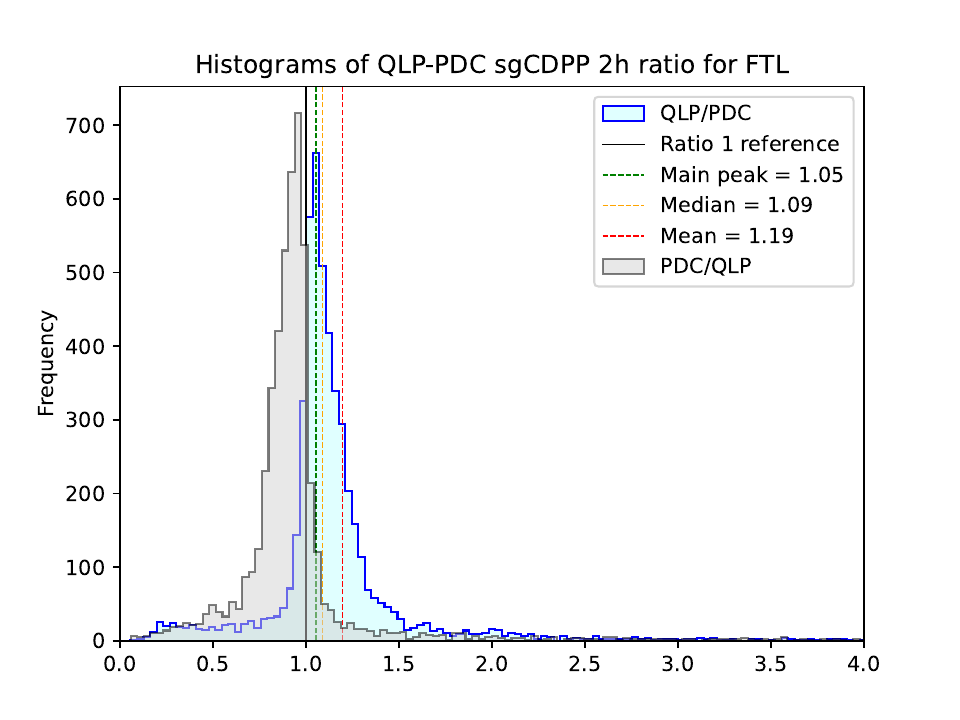}
\caption{Left column: sgCDPP measurements in ppm with time scales of 0.5 hours (top panel), 1 h (middle), and 2 h  (bottom) for an example sample of 5,000 cases in Sector 56. The top legends show various statistics for the ratio of the sgCDPP of QLP in the y-axis, over the sgCDPP of PDC in the x-axis; the number of cases with that ratio being above 1 and the corresponding percentage; a goodness-of-fit for the green, dashed line fitted to the data and the its fitted slope; and the number of cases used, and those excluded for having extreme ratios above 20. Right column: histograms of the sgCDPP ratios from QLP over those from PDC (in blue) and the inverse ratios (in gray). These panels also include a ratio 1 reference, vertical, solid line (in black), and vertical, dashed lines for the peak (green), median (orange), and mean (red) of the blue distribution.}
\label{fig:qlp_pdc_comparisons}
\end{figure*}

\section{Examples of PDC versus QLP comparisons}
\label{app:qlp_pdc_comparisons}

As discussed in Section~\ref{sec:flexible_targets}, using code from the branch \textsf{feature/flexible\_targets} of \textsf{SysCoCoPy}, we statistically compare sgCDPPs of PDC and QLP systematics-corrected light curves for a sample of 5,000 cases in Sector 56.~The latter was the first sector with available both 200-second cadence FFI data and QLP systematics-corrected products. For these QLP products, a new detrending algorithm utilizing quaternion data to remove short-timescale instrumental systematics is applied and intermediate data products containing only systematics-removed flux are included.~See results in Figure~\ref{fig:qlp_pdc_comparisons}, with a different time scale in each row (0.5, 1, and 2 hours, in descending order). In the left column, note the regions in which the sgCDPP of QLP is larger than that of PDC (i.e., above the ratio 1 reference line in blue) are more densely populated, as shown by the corresponding histograms in the right column; and most of the cases below the ratio 1 line have lower sgCDPP values.

\bibliography{SysCoCo}{}
\bibliographystyle{aasjournal}



\end{document}